\documentclass[journal,twoside,web]{ieeecolor}
\usepackage{generic}
\usepackage{cite}
\usepackage{amsmath,amssymb,amsfonts}
\usepackage{algorithmic}
\usepackage{graphicx}
\usepackage{algorithm,algorithmic}
\usepackage{hyperref}
\usepackage{textcomp}
\usepackage{tabularx}
\usepackage{overpic}
\usepackage{mathtools}
\usepackage{adjustbox}

\usepackage{orcidlink} %
\hypersetup{hidelinks} 

\def\BibTeX{{\rm B\kern-.05em{\sc i\kern-.025em b}\kern-.08em
    T\kern-.1667em\lower.7ex\hbox{E}\kern-.125emX}}
\begin{document}
\title{PTransIPs: Identification of phosphorylation sites enhanced by protein PLM embeddings}
\author{
Ziyang Xu$^{\orcidlink{0009-0008-5334-1155}}$, \IEEEmembership{Student Member, IEEE},
Haitian Zhong$^{\orcidlink{0009-0002-4741-3411}}$, Bingrui He, Xueying Wang$^{\orcidlink{0000-0001-8979-3737}}$ and Tianchi Lu$^{\orcidlink{0000-0001-6111-2020}}$, \IEEEmembership{Member, IEEE}
\thanks{(Ziyang Xu and Haitian Zhong contributed equally to this work.)}
\thanks{(Corresponding authors: Xueying Wang \& Tianchi Lu)}
\thanks{Ziyang Xu is with the School of Mathematics and Statistics, Lanzhou University, 222 South Tianshui Road, Lanzhou 730000, China (e-mail: xuzy20@lzu.edu.cn). }
\thanks{Haitian Zhong is with the Cuiying Honors College, Lanzhou University, 222 South Tianshui Road, Lanzhou 730000, China (e-mail: zhonght20@lzu.edu.cn).}
\thanks{Bingrui He is with the Lanzhou Power Supply Company, State Grid Gansu Electric Power Co., Ltd., Lanzhou, 730000, China (e-mail: qianqizy1501@gmail.com).}
\thanks{Xueying Wang is with the Department of Computer Science, City University of Hong Kong, Tat Chee Avenue, Kowloon, Hong Kong SAR, and also with the Department of Computer Science, City University of Hong Kong (Dongguan), Dongguan, China (e-mail: xywang85-c@my.cityu.edu.hk).}
\thanks{Tianchi Lu is with the Department of Computer Science, City University of Hong Kong, Tat Chee Avenue, Kowloon, Hong Kong SAR, and also with Suzhou Xuying Technology Co., Ltd. 2 Houtang Road, Suzhou 215164, China, and also with the School of Mathematics and Statistics, Lanzhou University, 222 South Tianshui Road, Lanzhou 730000, China(e-mail:tianchilu4-c@my.cityu.edu.hk). }
}
\maketitle

\begin{abstract}

Phosphorylation is pivotal in numerous fundamental cellular processes and plays a significant role in the onset and progression of various diseases. The accurate identification of these phosphorylation sites is crucial for unraveling the molecular mechanisms within cells and during viral infections, potentially leading to the discovery of novel therapeutic targets. In this study, we develop PTransIPs, a new deep learning framework for the identification of phosphorylation sites. Independent testing results demonstrate that PTransIPs outperforms existing state-of-the-art (SOTA) methods, achieving AUCs of 0.9232 and 0.9660 for the identification of phosphorylated S/T and Y sites, respectively. PTransIPs contributes from three aspects. 1) PTransIPs is the first to apply protein pre-trained language model (PLM) embeddings to this task. It utilizes ProtTrans and EMBER2 to extract sequence and structure embeddings, respectively, as additional inputs into the model, effectively addressing issues of dataset size and overfitting, thus enhancing model performance; 2) PTransIPs is based on Transformer architecture, optimized through the integration of convolutional neural networks and TIM loss function, providing practical insights for model design and training; 3) The encoding of amino acids in PTransIPs enables it to serve as a universal framework for other peptide bioactivity tasks, with its excellent performance shown in extended experiments of this paper. Our code, data and models are publicly available at \url{https://github.com/StatXzy7/PTransIPs}.

\end{abstract}

\begin{IEEEkeywords}
Phosphorylation sites, protein pre-trained language model, CNN, Transformer\\
\end{IEEEkeywords}

\section{Introduction}
\label{sec:introduction}



\IEEEPARstart{P}{hosphorylation}, a crucial post-translational modification process, plays a pivotal role in numerous fundamental cellular processes \cite{trewavas1976ref7,oliveira2012ref8}. This modification alters the structure and function of protein molecules by attaching phosphate groups to them. Phosphorylation significantly contributes to cell signal transduction, regulation of gene expression, control of the cell cycle, and the onset and progression of various diseases\cite{graves1999protein,mootha2004erralpha,lew1996regulatory,klann2020factorref4}. For example, the SARS-CoV-2 virus has had a substantial impact on human health and the global socioeconomic since its emergence in 2019\cite{barnes2020sars,wolf2023molecular,tutsoy2021unknown,tutsoy2022linear}. Studies have shown that the phosphorylation state of its nucleocapsid protein affects the virus's activity, suggesting that phosphatases could be potential drug targets \cite{acter2020SARS-CoV-2, tugaeva2021ref10, eisenreichova2022ref11, patel2021ref12}. Therefore, a deeper understanding of phosphorylation holds immense value for biomedical research.

Nowadays, high-throughput sequencing technologies can provide us with a large amount of accurate phosphorylation site data \cite{huang2019highref13, hekman2020actionable}, but expensive equipment and experimental costs remain a challenge for many laboratories. Building reliable phosphorylation site identification models through computational methods can guide the design of experimental schemes and the analysis of results, reducing sequencing costs, and thus is of significant importance.

To date, several predictors for identifying phosphorylation sites have been proposed. Traditional machine learning methods have achieved commendable results in the past. For instance, PhosPred-RF employs a combination of various features with a random forest algorithm\cite{wei2017phospredPhosPred-RF}, Quokka utilizes sequence scoring functions combined with logistic regression algorithms\cite{li2018quokka}, and GPS 5.0 adopts position weight and scoring matrix combined with logistic regression algorithms for predicting phosphorylation sites\cite{wang2020gps}. These algorithms mainly rely on manually designed feature extraction methods, thus possessing significant limitations. In recent years, several deep learning-based models have completed this task with higher performance\cite{luo2019deepphos,wang2019capsule}. For example, MusiteDeep uses convolutional neural networks (CNNs) with a two-dimensional attention mechanism to predict phosphorylation sites\cite{wang2017musitedeep, wang2020musitedeep}. 
DeepPSP is a deep neural network based on global-local information for the prediction of phosphorylation sites\cite{guo2021deeppsp}. Lv et al. introduced DeepIPs, constructing a CNN-LSTM framework for prediction\cite{lv2021deepips}. Wang et al. used feature learning through differential evolution combined with a multi-head attention mechanism for prediction, achieving an AUC of over 90\%\cite{wang2023DE-MHAIPs}.


However, given that phosphorylation is a post-translational modification process on protein molecules, these models learnt from limited samples may not adequately capture the characteristics of proteins, leading to insufficient generalization capabilities of the model. Therefore, to further enhance predictive performance, it is necessary to explore methods for extracting additional information from samples.
The outstanding performance of pre-trained language models (PLMs) in content generation in recent years has inspired us\cite{devlin2019bert}. These models are pre-trained on large-scale unlabeled corpora, learning contextual word representations, which makes them highly effective as universal semantic features. For instance, protein PLMs have achieved significant progress in the field of protein structure prediction\cite{jumper2021highlyAlphaFold2,lin2022language,ProtTrans,weissenow2022EMBER2}. Thus, the embeddings generated by inputting sequences into these models may contain a large amount of additional information we need.

\begin{figure*}[ht]
    \centering
    \includegraphics[width=1\textwidth]{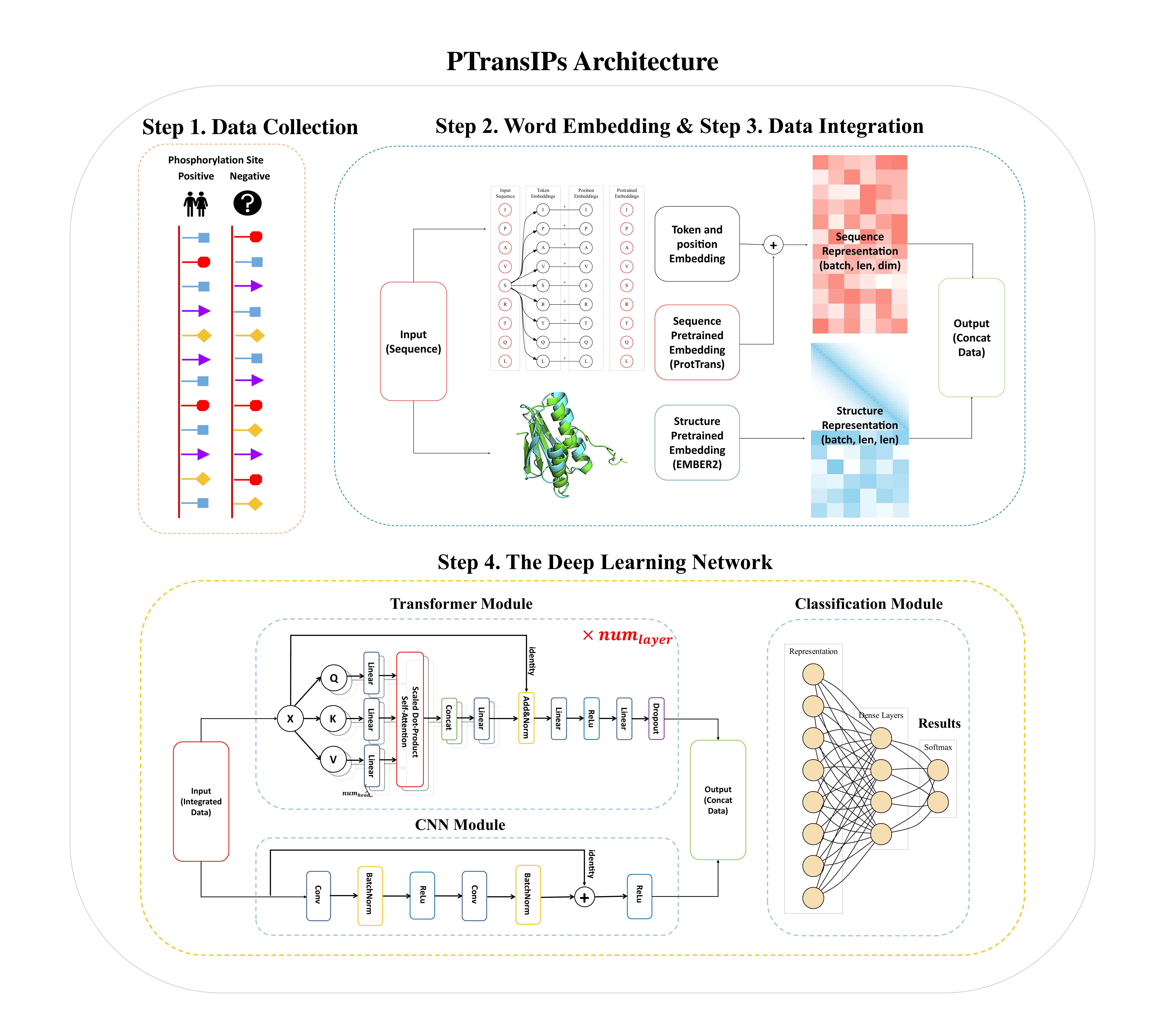}
    \caption{PTransIPs architecture. The figure illustrates the steps of the PTransIPs model for identifying SARS-CoV-2 phosphorylation sites. It starts with data collection (Step 1) where S/T and Y phosphorylation sites dataset is gathered. Next, in the word embedding phase (Step 2), a unique 1024-dimensional vector representation for each amino acid type in the sequence is constructed. Data integration (Step 3) combines these embeddings to enhance the representational capacity of input data. The integrated data are then processed in parallel by a CNN with residual connections and a Transformer based on multi-head attention in the deep learning network phase (Step 4). The outputs of the two models are then connected to a fully connected layer classifier to predict the phosphorylation sites.}
    \label{fig:architecture}
\end{figure*}

In our study, we propose a novel deep learning model, PTransIPs, for the identification of phosphorylation sites. As illustrated in Figure \ref{fig:architecture}, the model treats amino acids in protein sequences as words, extracting unique encodings based on the types of amino acids and their positions in the sequence. Embeddings generated from protein PLMs are also considered a form of encoding input into the model. PTransIPs is further trained on a combined CNN and Transformer model, ultimately outputting classification results through a fully connected layer. To validate the performance of PTransIPs, we conduct independent testing after model training. The results reveal that PTransIPs achieves AUCs of 0.9232 and 0.9660 for identifying phosphorylated S/T and Y sites respectively, surpassing existing state-of-the-art (SOTA) methods. Furthermore, we conduct ablation studies to confirm the contribution of pre-trained model embeddings to prediction efficacy. To test the model's generalizability, we extend its application beyond phosphorylation to other biological activity classification tasks, achieving optimal results on certain metrics. To facilitate usage, we have made our code and data publicly accessible at \url{https://github.com/StatXzy7/PTransIPs}.

\section{Materials and methods}\label{sec2}

\subsection{Datasets}\label{subsec1}

The primary raw data used in this article includes experimentally verified phosphorylation sites extracted from human A549 cells infected with COVID-19, collected from the literature \cite{stukalov2021multi}. We employed the same preprocessing method as Lv et al. \cite{lv2021deepips} to generate a dataset suitable for model training, ensuring fairness in comparison. The steps are as follows. First, the CD-HIT tool \cite{li2006cdhit} was used to discard data with more than 30\% protein sequence similarity, to limit sequence redundancy. Second, the retained sequences were segmented into 33-residue fragments, with S/T or Y located at the center. A phosphorylated S/T or Y at the midpoint of a fragment categorizes it as a positive sample, otherwise, it is a negative sample. Third, a subset of non-redundant negative samples was randomly selected to match the quantity of positive samples, to balance the positive and negative data. Ultimately, the resulting dataset comprises 10,774 S/T site samples and 204 Y site samples, with a balanced number of positive and negative samples for each type. The samples were then divided into non-overlapping training and testing sets, maintaining an 8:2 distribution \cite{wei2020computational}. Table \ref{dataset} provides details on the composition of the dataset.


\begin{table}[ht]
\centering
\caption{Distribution of Positive and Negative Samples for S/T and Y Phosphorylation Sites in Training and Test Datasets}
\label{dataset}
\begin{tabularx}{\columnwidth}{p{3cm} p{2cm} X X}
\hline Datasets & Types & $\mathrm{S} / \mathrm{T}$ & $\mathrm{Y}$ \\
\hline Training set & Positive & 4308 & 81 \\
& Negative & 4308 & 81 \\
Independent test set & Positive & 1079 & 21 \\
& Negative & 1079 & 21 \\
\hline
\end{tabularx}
\end{table}

\subsection{Token and position embedding}




Our model employs an embedding strategy inspired by BERT\cite{devlin2019bert}, incorporating both Token embedding and Position embedding components. 

For Token embedding, we construct unique vector representations for different types of amino acids in the sequence, mapping each amino acid to a vector of a fixed embedding dimension (dim=1024). For Position embedding, we similarly create a unique vector representation for each position based on the sequence length (length=33) and embedding dimension (dim=1024). We implement these two embeddings using 'nn.Embedding' function in Python package 'Pytorch'\cite{paszke2019pytorch}.

Specifically, for an amino acid $x$ in the position $i$ of a sequence, its embedding can be calculated as:
\begin{equation}
Emb(x, i) = LN(Emb_{token}(x)+Emb_{pos}(i))
\end{equation}
where $x$ represents the type of amino acid, and $i$ represents its position in the sequence. $Emb_{token}(x)$ is the token embedding of $x$, and $Emb_{pos}(i)$ is the position embedding for position $i$. $LN$ denotes the layer normalization operation, which enhances training stability. The embedding dimension is chosen to be 1024 to provide sufficient capacity to capture sequence features.

\subsection{pre-trained embeddings for sequence and structure}




To extract additional features from protein sequences, we utilized two protein pre-trained language models: ProtTrans and EMBER2. To generate comprehensive sequence embeddings, we employ ProtTrans\cite{ProtTrans}, a transformative self-supervised learning model pioneered by Elnaggar et al. in our study. Using ProtTrans pre-trained model, we encoded each amino acid in the sequence into a unique 1024-dim vector, whose dimension is the same as the token and position embedding.
Also, we utilize EMBER2\cite{weissenow2022EMBER2}, an advanced model adept at protein structure prediction to obtain additional structure information of our sequence. For each sequence, we generate the contact matrix, distance matrix of average and distance matrix of mode. The dimension of each embedding is $len \times len$, where the $len = 33$ here represents the length of the sequence in our dataset.

By incorporating these two protein pre-trained language models, we are able to generate robust and meaningful representations of amino acid sequences, which provides enhanced information as embeddings for the coming training process.

\subsection{The architecture of PTransIPs}

Our study introduces PTransIPs, a novel methodology designed to identify SARS-CoV-2 phosphorylation sites. A visual representation of its workflow is provided in Figure \ref{fig:architecture}.

The PTransIPs procedure includes:

Step1. Data collection. Dataset of S/T and Y phosphorylation sites are collected from \cite{lv2021deepips}. These data include amino acid sequences and corresponding labels.

Step2. Word embedding. We use the token and position embedding method to construct a unique 1024-dimensional vector representation for each amino acid type in the sequence. Also, we utilize two protein pre-trained language models to obtain their embeddings as additional information.

Step3. Data integration. We combine the embeddings obtained in the previous step using addition and concatenation methods to enhance the representational capacity of input data.

Step4. The deep learning network. The integrated data are fed in parallel into a CNN with residual connections and a Transformer based on multi-head attention. The results obtained are then connected to a fully connected layer classifier to predict the SARS-CoV-2 phosphorylation sites.

Step5. Performance evaluation. To assess the efficacy of the model spanning Step1 through Step4, we employ a 5-fold cross-validation approach. Metrics such as the Area Under the ROC Curve (AUC), Accuracy (ACC), Sensitivity (SEN), Specificity (SPEC), and Matthews Correlation Coefficient (MCC) are selected for examination of prediction results. Following the identification of the most effective prior model, we evaluate its performance on the independent test data.

\subsubsection{Data integration}\label{Data integration}

To this step, we have obtained self-embeddings based on token and position embedding $Emb\in \mathbb{R}^{batch_{size} \times seq_{len} \times dim_{emb} = 1024}$, sequence embeddings generated by the pre-trained protein model ProtTrans $preEmb_{seq}\in \mathbb{R}^{batch_{size} \times seq_{len} \times dim_{emb} = 1024}$, and structural embeddings generated by the pre-trained protein model EMBER2 $preEmb_{str}\in \mathbb{R}^{batch_{size} \times seq_{len} \times dim_{str} = 256}$. To combine these embeddings, we first add up the self-embeddings and the pre-trained sequence embeddings. Next, we concatenate the aggregated embeddings with the structural embeddings along their last dimension.
Through these steps, we create integrated data $X \in \mathbb{R}^{batch_{size} \times seq_{len} \times dim = 1280}$ that captures both the sequential and structural information of protein sequences, thereby enhancing the representation capability of the input data for subsequent analysis and prediction tasks.
\begin{equation}
    X = \text{Concat}( (Emb{seq}+preEmb_{seq}) , preEmb_{str} )
\end{equation}

\subsubsection{The Transformer module}\label{Transformer}




Following the step of concatenating embedding vectors to form the integrated data $X \in \mathbb{R}^{batch_{size} \times seq_{len} \times dim}$, we incorporate the Transformer architecture \cite{vaswani2017attentionisallyouneed} into our training process. The core of Transformer is the multi-head attention mechanism \cite{vaswani2017attentionisallyouneed,bahdanau2014neural,kim2017structured,parikh2016decomposable}, enabling comprehensive feature extraction. Given input vectors denoted as query ($Q$), key ($K$), and value ($V$), the scaled dot-product attention is computed as follows:
\begin{equation}
\text{Attention}(Q, K, V) = \text{softmax}\left(\frac{QK^T}{\sqrt{d_k}}\right)V
\end{equation}
where $d_k$ is the dimension of the key vectors. The scaling by $\sqrt{d_k}$ serves as a normalization factor, ensuring that an increase in dimensions does not lead to a significant escalation in the dot product.
The multi-head attention is computed by linearly transforming the input vectors $Q$, $K$, $V$ $h$ times (where $h$ represents the number of attention heads), applying the scaled dot-product attention to each of these transformed vectors, and then concatenating and linearly transforming the results as follows:
\begin{equation}\label{eq:multihead_attention}
\begin{split}
\text{MultiHead}(Q, K, V) = \text{Concat}(\text{head}_1, \ldots, \text{head}_h)W_O \\
\text{where} \; \text{head}_i = \text{Attention}({Q_i}, {K_i}, {V_i})
\end{split}
\end{equation}
For attention head $i$, $W_{Q_i}$, $W_{K_i}$, and $W_{V_i}$ are the weight matrices for the $Q$, $K$, $V$ vectors, respectively, allowing the computation of $Q_i = X W_{Q_i}$, $K_i = X W_{K_i}$, and $V_i = X W_{V_i}$. $W_O$ is the output weight matrix.


\subsubsection{The CNN module}\label{CNN}
In parallel with the Transformer module, we also integrate a Convolutional Neural Network (CNN) module \cite{lecun1989backpropagation,krizhevsky2012imagenetcnn,simonyan2015vgg} as part of our training framework. Inspired by ResNet\cite{he2016deep}, the core component of our CNN module is a residual block that consists of two 1D convolutional layers. For the input integrated data $X \in \mathbb{R}^{batch_{size} \times seq_{len} \times dim}$, the residual block is applied directly to $X$, with the computation formula as follows:
\begin{equation}
\begin{aligned}
& F_1(X_l) = \text{Conv1D}(W_1, \text{ReLU}(\text{BN}(X_l))) + b_1 \\
& F_2(X_l) = \text{Conv1D}(W_2, \text{ReLU}(\text{BN}(F_1(X_l)))) + b_2 \\
& X_{l+1} = X_l + F_2(X_l) 
\end{aligned}
\end{equation}
Where $X_l$ represents the input to the $(l+1)^{th}$ residual block, and $X_{l+1}$ represents the output of that residual block. $\text{Conv1D}$ represents the 1D convolution operation, $W_i$ and $b_i$ are the weights and biases of the $i^{th}$ layer within the residual block, $\text{ReLU}$ is the Rectified Linear Unit activation function, and $\text{BN}$ denotes the Batch Normalization operation. 

\subsubsection{The TIM loss function}



The training process of PTransIPs is inspired by the Transductive Information Maximization (TIM) loss function\cite{boudiaf2020transductive}, which is a combination of the traditional cross-entropy loss and empirically weighted mutual information. Given our labeled dataset and the supervised learning task, we utilize a variant of the TIM Loss function that calculates all losses solely on the training set. Specifically, the empirical mutual information between the data X (amino acid sequences) and their corresponding labels Y (indicating whether it is a phosphorylation site or not) is divided into two main components. The first component is the empirical conditional entropy of the labels, denoted as $\widehat{\mathcal{H}}\left(Y \mid X \right)$. The second component is the empirical marginal entropy of the labels, denoted as $\widehat{\mathcal{H}}\left(Y \right)$. Additionally, the cross-entropy loss between the labels and the data, denoted as $\mathrm{CE}$, should also be considered to optimize for binary classification. The calculation for these three components are as follows:
\begin{equation}
    \begin{aligned}
        &  \widehat{\mathcal{H}}(Y) := -\sum_{k=1}^K \widehat{p}_k \log \widehat{p}_k  \\
        & \widehat{\mathcal{H}}(Y \mid X) := - \frac{1}{|X|} \sum_{i \in X} \sum_{k=1}^K p_{ik} \log(p_{ik}) \\
        &  \mathrm{CE}:=-\frac{1}{|{X}|} \sum_{i \in {X}} \sum_{k=1}^K y_{i k} \log \left(p_{i k}\right) \\
    \end{aligned}
    \label{TIM Loss equation terms}
\end{equation}
Where $|X|$ is the size of the dataset, $i$ indexes the dataset $X$, and $k$ indexes the label categories. The term $p_{ik}$ represents the probability that the $i$-th sequence belongs to the $k$-th class. $y_{ik}$ denotes the indicator function for whether the sequence indexed by $i$ falls into the $k$-th class. We set $K = 2$, as the task for this study is binary classification.


The final loss function for PTransIPs is defined as:
\begin{equation}
    \widehat{\mathcal{L}}(X ; Y) :=  \lambda \mathrm{CE} - \widehat{\mathcal{H}}(Y) + \alpha \widehat{\mathcal{H}}(Y \mid X) 
\end{equation}
Where $\alpha$ and $\lambda$ are hyperparameters that determine the rate of convergence for each term in the loss function. Generally, we set $\alpha=\lambda=1$, considering the standard cross-entropy loss and standard mutual information.

\subsection{Hyperparameter setting}


The PTransIPs model is implemented in Python using PyTorch. For the Transformer module, the number of multi-head attention layers is set to 6, and the number of attention heads to 8. For the CNN module, the input and output channels are set to $\mathrm{c}_{in} = \mathrm{c}_{out} = 1280$, with a kernel size of $k=5$, stride of $p=1$, and padding of $2$. The model uses the Adam optimizer, with an initial learning rate of 0.00001, and is trained for 100 epochs. All computational experiments and results were conducted in the environment: Python 3.9, GPU RTX $3090(24 \mathrm{~GB}) \times 1$, CPU Intel® Xeon® Gold 6330, and 80GB RAM.

\subsection{Performance evaluation}

For the assessment of our deep learning model's capabilities, we turn to several evaluation metrics, notably ACC, SPEC, SEN and MCC. These metrics are defined as:

\begin{equation}
A C C=\frac{T P+T N}{T P+F P+T N+F N}
\end{equation}

\begin{equation}
S E N=\frac{T P}{T P+F N}
\end{equation}

\begin{equation}
S P E C=\frac{T N}{T N+F P}
\end{equation}

\begin{equation}
\resizebox{.9\hsize}{!}{$MCC = \frac{TP \times TN - FP \times FN}{\sqrt{(TP + FP) \times (TP + FN) \times (TN + FP) \times (TN + FN)}}$}
\end{equation}

TP represents the accurate identification number of positive phosphorylation sites. TN represents the accurate identification number of negative phosphorylation sites. FP represents the incorrect identification number of positive phosphorylation sites. FN represents the incorrect identification number of negative phosphorylation sites.

In addition to the metrics previously described, we can further assess classification performance utilizing the Receiver Operating Characteristic (ROC) and Precision-Recall (PR) curves. The model's efficacy is quantified by the Area Under the ROC Curve (AUC) and the Area Under the Precision-Recall Curve (AUPR).

\section{Results}\label{sec3}               

\subsection{Evaluating the contribution of pre-trained model embedding to results}

\begin{figure*}[ht]
  \begin{minipage}[b]{0.45\textwidth}
    \begin{overpic}[width=\textwidth]{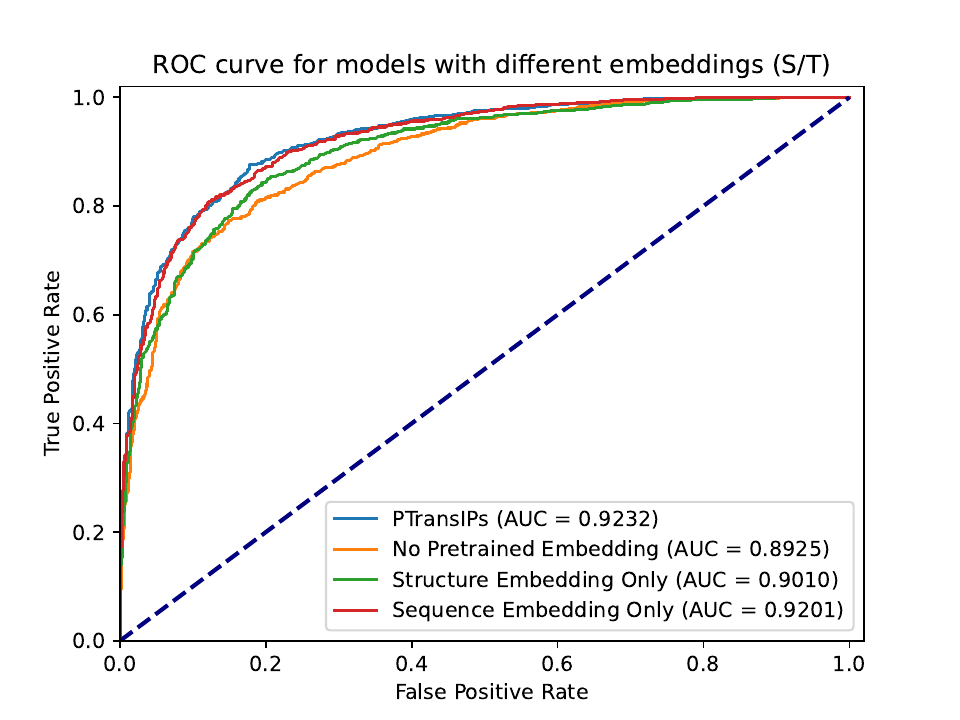} 
      \put(2,70){\large\textbf{A}}
    \end{overpic}

    \vspace{2mm}

    \begin{overpic}[width=\textwidth]{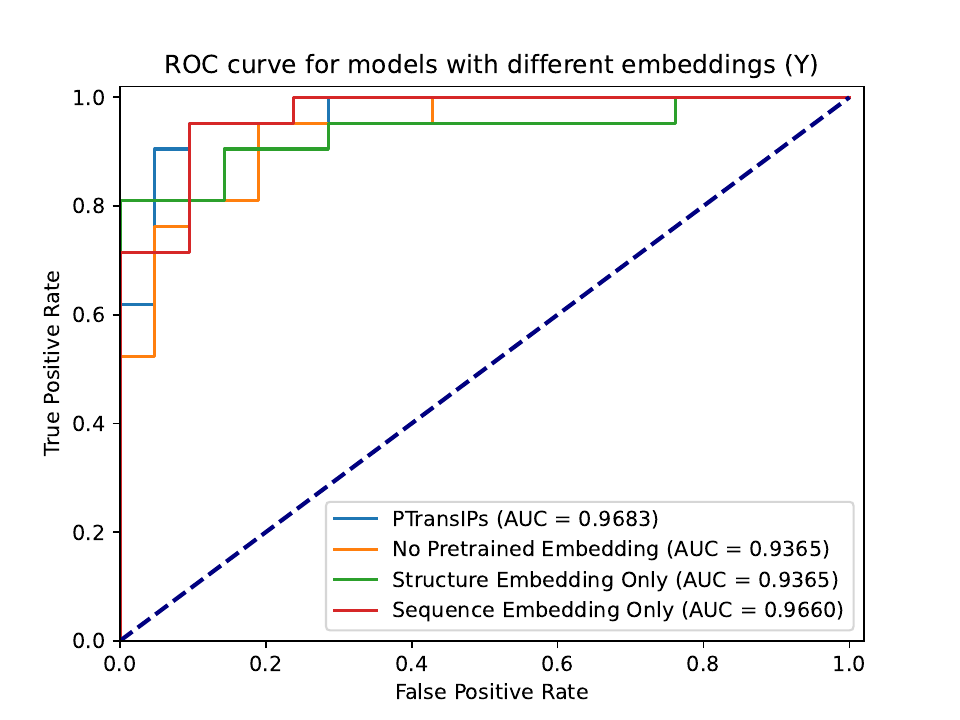}
      \put(2,70){\large\textbf{C}}
    \end{overpic}
  \end{minipage}
  \hspace{2mm}
  \begin{minipage}[b]{0.45\textwidth}
    \begin{overpic}[width=\textwidth]{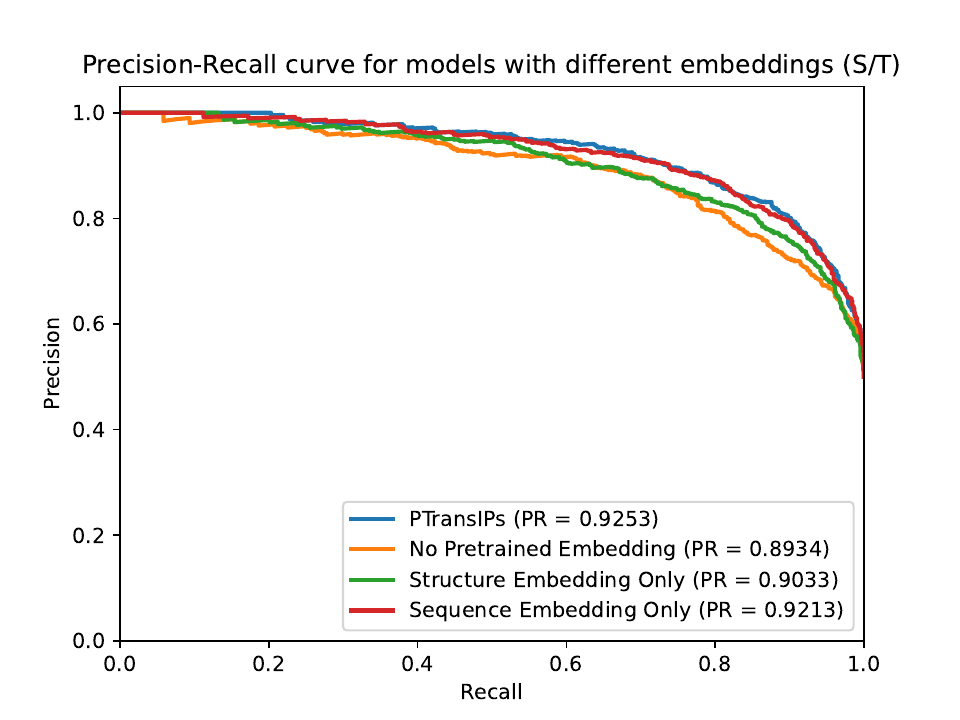}
      \put(2,70){\large\textbf{B}}
    \end{overpic}

    \vspace{2mm}

    \begin{overpic}[width=\textwidth]{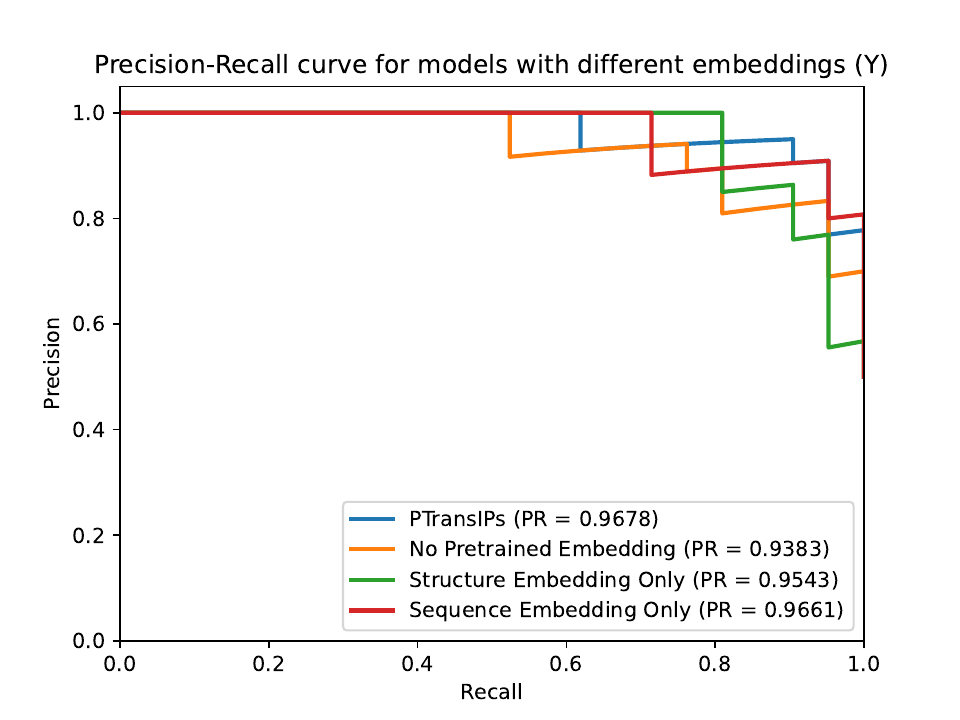}
      \put(2,70){\large\textbf{D}}
    \end{overpic}
    
  \end{minipage}
  \caption{ROC and PR curves for phosphorylation site identification for ablation study on pre-trained embedding.
This figure shows the comparison of ROC and PR curves among PTransIPs, the model using only sequence pre-trained embedding, the model using only structure pre-trained embedding, and the model without any pre-trained embeddings. (A–B) show the ROC and PR curves for the S/T dataset, while (C–D) show the same curves for the Y dataset.}
  \label{fig:ablation on pre-trained}
\end{figure*}

In this section, we conduct an ablation study aimed at evaluating the impact of using pre-trained embeddings on PTransIPs. For this purpose, we designed four models: PTransIPs itself, the model using only sequence pre-trained embeddings, the model using only structure pre-trained embeddings, and the model not using any pre-trained embeddings. We train these four models separately on the training set in Section \ref{subsec1}, and test them on the corresponding independent test set. The ROC and PR curves for these methods are plotted in Figure \ref{fig:ablation on pre-trained}, and all evaluation metrics are shown in Table \ref{Table:ablation performance}.

\begin{table*}
\centering
\caption{Independent Testing Performance Comparison among models enhanced by pre-trained embeddings or not for S/T and Y Sites}
\label{Table:ablation performance}
\begin{tabularx}{\textwidth}{X p{4cm} XXXXX}
\hline Residue type & Method & ACC & SEN & SPEC & MCC & AUC \\
\hline S/T & PTransIPs & \textbf{0.8438} & \textbf{0.8554} & \textbf{0.8323} & \textbf{0.6879} & \textbf{0.9232} \\
& Sequence Embedding Only & 0.8336 & 0.8378 & 0.8295 & 0.6673 & 0.9201 \\
& Structure Embedding Only & 0.8253 & 0.8350 & 0.8156 & 0.6507 & 0.9010 \\
& No pre-trained Embedding & 0.8072 & 0.8063 & 0.8082 & 0.6145 & 0.8925 \\
\hline
Y & PTransIPs & \textbf{0.9286} & \textbf{0.9524} & 0.9048 & \textbf{0.8581} & \textbf{0.9683} \\
& Sequence Embedding Only & 0.9286 & 0.9048 & \textbf{0.9524} & 0.8581 & 0.9660\\
& Structure Embedding Only & 0.8571 & 0.8095 & 0.9048 & 0.7175 & 0.9365 \\
& No pre-trained Embedding & 0.8810 & 0.8571 & 0.9048 & 0.7628 & 0.9365 \\
\hline
\end{tabularx}
\end{table*}


Overall, PTransIPs outperforms other models on nearly all evaluation metrics, demonstrating that the use of pre-trained embeddings can enhance the overall performance of identification. Specifically, for S/T sites, PTransIPs performs better across all evaluation metrics. For Y sites, PTransIPs and the model using only sequence pre-trained embeddings have almost identical performances, still superior to models without any pre-trained embeddings.

It is noteworthy that for the identification of both types of sites, the pre-trained embeddings of sequences significantly enhance prediction performance, while the contribution of structural information is relatively minor. We believe there are two main reasons for this: firstly, the original dimension of sequence embeddings is approximately 10 times that of structure embeddings, containing more information; secondly, the positive and negative sequences in our dataset are relatively similar, which makes the predicted protein structure information less capable of distinguishing their differences.

\subsection{Training with the TIM loss function improves the performance of PTransIPs}

\begin{figure*}[ht]
  \begin{minipage}[b]{0.45\textwidth}
    \begin{overpic}[width=\textwidth]{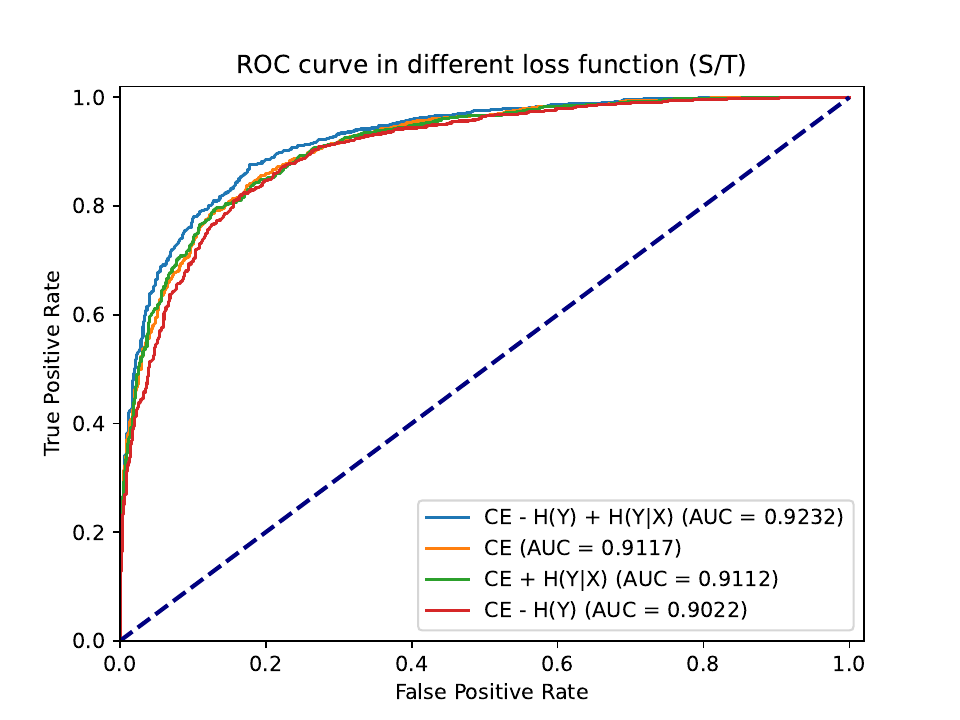} 
      \put(2,70){\large\textbf{A}}
    \end{overpic}

    \vspace{2mm}

    \begin{overpic}[width=\textwidth]{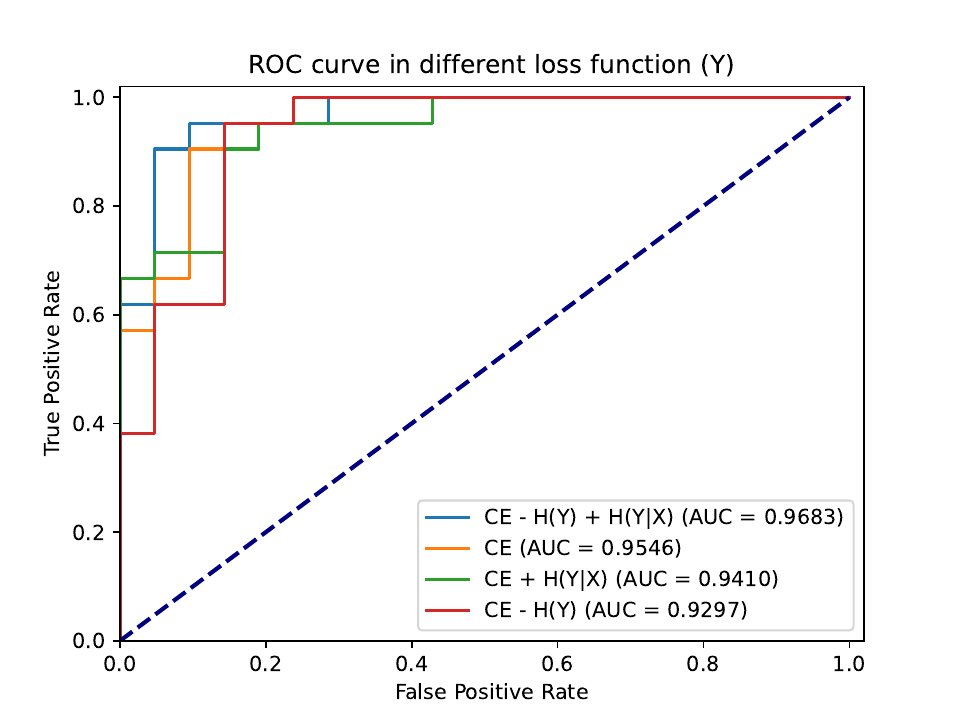}
      \put(2,70){\large\textbf{C}}
    \end{overpic}
  \end{minipage}
  \hspace{2mm}
  \begin{minipage}[b]{0.45\textwidth}
    \begin{overpic}[width=\textwidth]{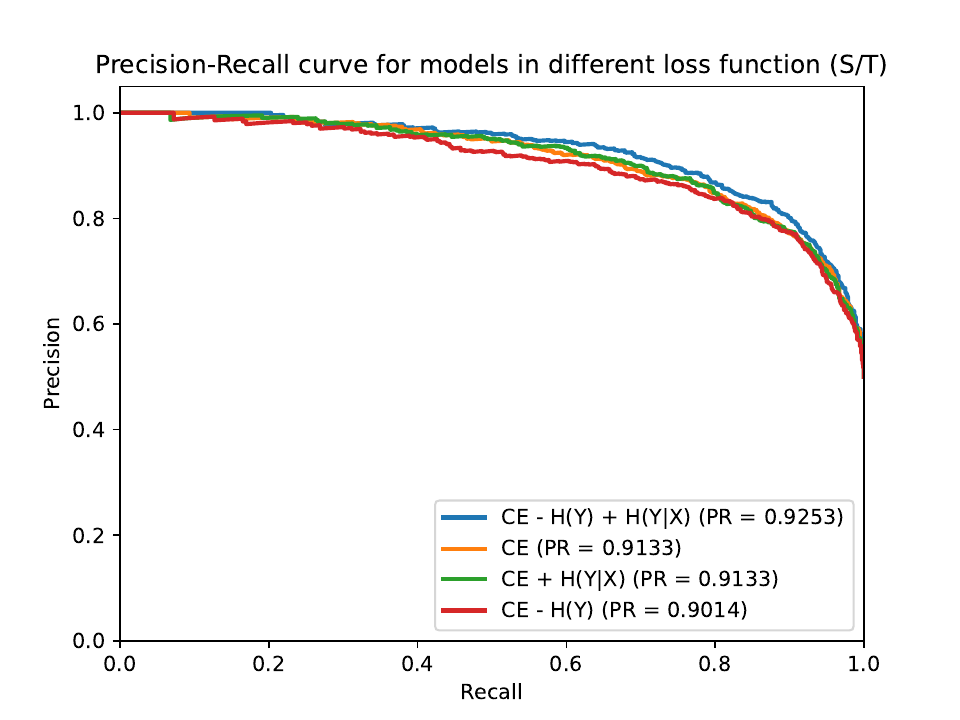}
      \put(2,70){\large\textbf{B}}
    \end{overpic}

    \vspace{2mm}

    \begin{overpic}[width=\textwidth]{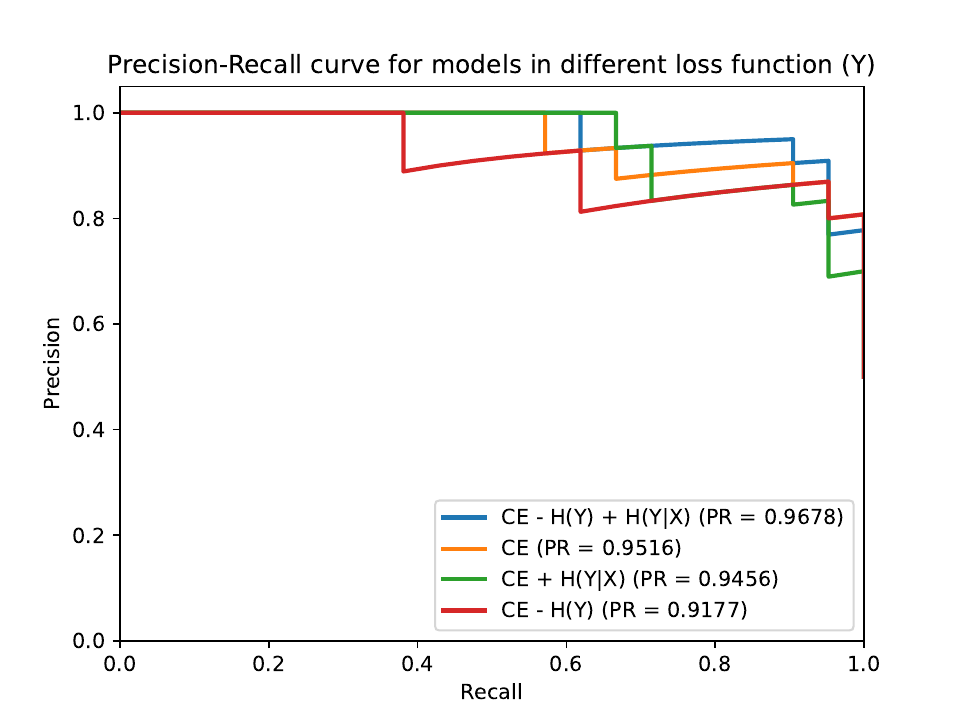}
      \put(2,70){\large\textbf{D}}
    \end{overpic}
    
  \end{minipage}
  \caption{ROC and PR Curves for Phosphorylation Site Identification in the Ablation Study of the Three Terms of the TIM Loss Function.
This figure shows the comparison of ROC and PR curves between the complete TIM Loss function used by PTransIPs $ CE - \widehat{\mathcal{H}}\left(Y\right) +\widehat{\mathcal{H}}\left(Y \mid X\right)$, the original cross-entropy loss $CE$, and the loss functions with either Marginal Entropy $\widehat{\mathcal{H}}\left(Y \right)$ or Conditional Entropy $\widehat{\mathcal{H}}\left(Y \mid X \right)$ removed. (A-B) show the ROC and PR curves for the S/T dataset, while (C-D) show the same curves for the Y dataset.}
  \label{fig:ablation on loss function}
\end{figure*}

In this section, we evaluate the impact of training with the TIM Loss function on the performance of PTransIPs. For this purpose, we conduct an ablation study on the contribution of each term in the TIM Loss function to the overall loss. The ROC and PR curves obtained from training and testing with these methods are shown in Figure \ref{fig:ablation on loss function}, and all evaluation metrics are shown in Table \ref{Table Performance evaluation on loss function}. Here, the notation for each term follows that of Equation \ref{TIM Loss equation terms}: CE: Cross-Entropy, $\widehat{\mathcal{H}}\left(Y \right)$: Marginal Entropy, $\widehat{\mathcal{H}}\left(Y \mid X \right)$: Conditional Entropy.

We observe that using the complete TIM Loss with all three terms consistently outperforms any other Loss Function. Specifically, removing the label marginal entropy $\widehat{\mathcal{H}}\left(Y \right)$ significantly reduces model performance. This phenomenon can be theoretically explained by the fact that optimizing solely on the conditional entropy term $\widehat{\mathcal{H}}\left(Y \mid X \right)$ may lead to degenerate solutions that assign all data to a single category, resulting in performance degradation. This also underscores the importance of marginal entropy $\widehat{\mathcal{H}}\left(Y \right)$ as a regularization term for enhancing the model's generalization performance.

\begin{table*}
\centering
\caption{Independent Testing Performance Comparison among models trained with different loss functions for S/T and Y Sites}
\label{Table Performance evaluation on loss function}
\begin{tabularx}{\textwidth}{X p{5cm} XXXXX}
\hline Residue type & Loss function & ACC & SEN & SPEC & MCC & AUC \\
\hline S/T & $ CE - \widehat{\mathcal{H}}\left(Y\right) +\widehat{\mathcal{H}}\left(Y \mid X\right)$ & \textbf{0.8438} & \textbf{0.8554} & \textbf{0.8323} & \textbf{0.6879} & \textbf{0.9232} \\
& $ CE - \widehat{\mathcal{H}}\left(Y\right) $ & 0.8299 & 0.8462 & 0.8137 & 0.6602 & 0.9137 \\
& $ CE +\widehat{\mathcal{H}}\left(Y \mid X\right)$ & 0.8258 & 0.8443 & 0.8072 & 0.6520 & 0.9112 \\
& $ CE $ & 0.8234 & 0.8360 & 0.8109 & 0.6471 & 0.9117 \\
\hline
Y & $ CE - \widehat{\mathcal{H}}\left(Y\right) +\widehat{\mathcal{H}}\left(Y \mid X\right)$ & \textbf{0.9286} & \textbf{0.9524} & \textbf{0.9048} & \textbf{0.8581} & \textbf{0.9683} \\
& $ CE - \widehat{\mathcal{H}}\left(Y\right) $ & 0.8571 & 0.8571 & 0.8571 & 0.7143 & 0.9297\\
& $ CE +\widehat{\mathcal{H}}\left(Y \mid X\right)$ & 0.8571 & 0.8571 & 0.8571 & 0.7143 & 0.9410 \\
& CE & 0.8333 & 0.7619 & 0.9048 & 0.6736 & 0.9546 \\
\hline
\end{tabularx}
\end{table*}

\subsection{Visualizing the feature extraction process of PTransIPs with UMAP}

\begin{figure*}[ht]
    \centering
    \includegraphics[width=1\textwidth]{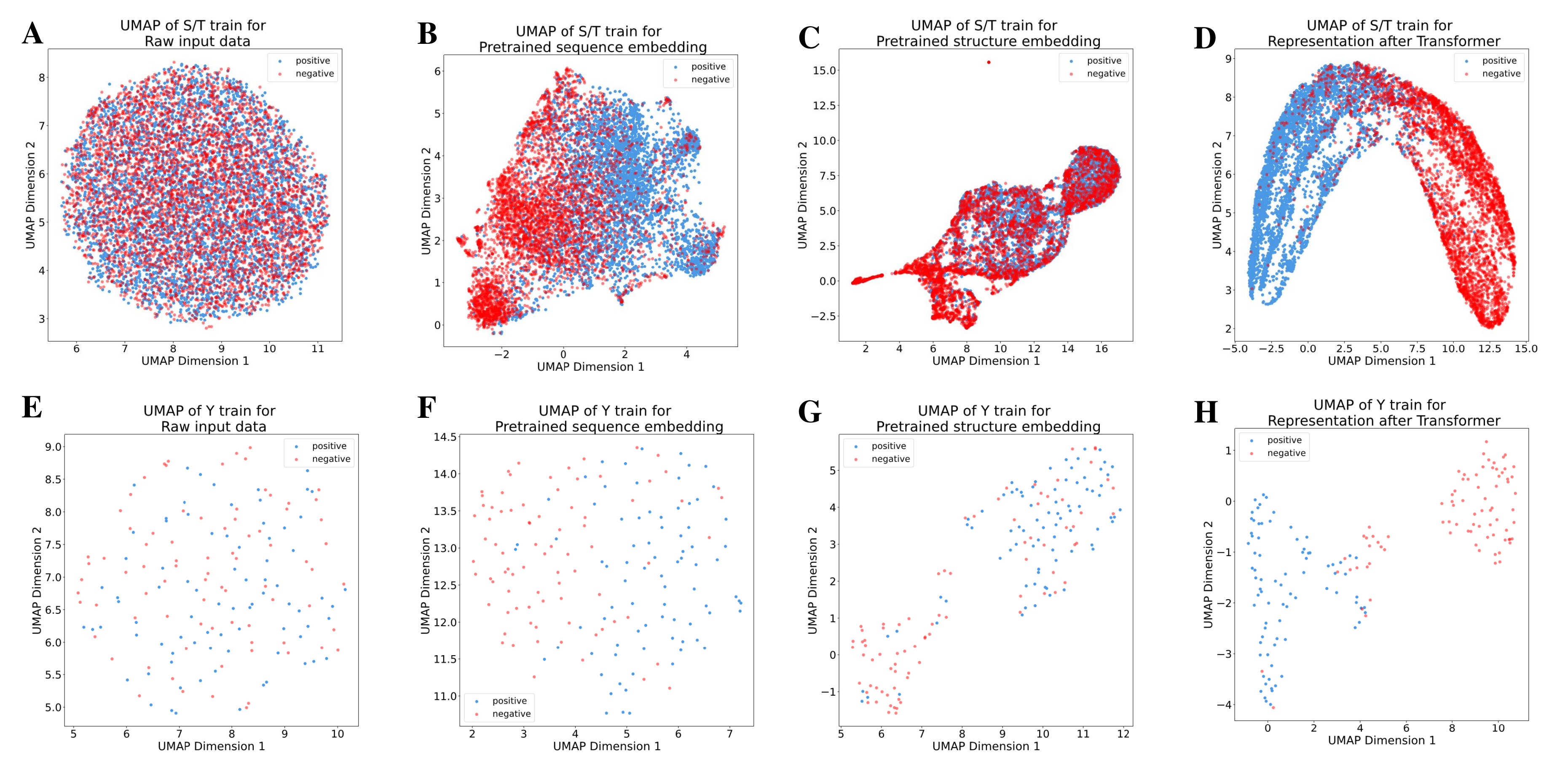}
    \caption{UMAP-based 2D Feature Space Distribution of Positive and Negative Samples for S/T and Y Training Sets.
  The figure shows the distribution of S/T and Y sites in the feature space generated by UMAP, based on the original features from input data (A, E), features from the pre-trained sequence model (B, F), features from the pre-trained structure model (C, G), and output features from the deep learning network (CNN and Transformer modules) (D, H). Blue and red dots represent positive and negative samples, respectively.}
    \label{fig:Visualization of PTransIPs training process}
\end{figure*}

To investigate our model's ability to distinguish phosphorylation sites, we visualize the features extracted by PTransIPs during different stages of training process using uniform manifold approximation and projection (UMAP)\cite{mcinnes2018umap, becht2019dimensionalityumap}. For both S/T and Y phosphorylation sites, the distinction between positive and negative samples was nebulous based on the raw input, and the data points appeared intermingled and lacked clear boundaries (Figure \ref{fig:Visualization of PTransIPs training process}A). However, for embeddings from both sequence and structure protein pre-trained models, the pre-trained sequence model demonstrated robust discriminatory power (Figure \ref{fig:Visualization of PTransIPs training process}B), and the pre-trained structural model also showed preliminary differentiation capability between the two types of samples (Figure \ref{fig:Visualization of PTransIPs training process}C), intuitively proving the utility of using pre-trained models for training. Progressing further, once data was processed via the combined features of CNN and Transformer layers, the demarcation between positive and negative samples became stark and apparent (Figure \ref{fig:Visualization of PTransIPs training process}D). These observations not only indicate that the features extracted by PTransIPs possess the capability to identify phosphorylation sites, but also intuitively demonstrate that the embeddings generated from protein PLMs can distinguish whether sequences are phosphorylated to some extent. The visualization validates the effectiveness of the PTransIPs architecture and training process.


In addition, we plot the corresponding UMAP figures for the test dataset of S/T and Y phosphorylation sites. These results are similar to those presented in the main text for the train dataset of S/T sites and can be found in supplementary material Figure S1-S3.

\subsection{Independent test of PTransIPs for phosphorylation site
identification}\label{Evaluation of PTransIPs}

To evaluate the performance of PTransIPs further, we compare it with five existing phosphorylation site identification tools using data from the independent test: DeepIPs\cite{lv2021deepips}, DE-MHAIPs\cite{wang2023DE-MHAIPs}, DeepPSP\cite{guo2021deeppsp}, MusiteDeep-2020\linebreak[1]\cite{wang2020musitedeep}, and MusiteDeep2017\cite{wang2017musitedeep}. For the sake of fairness in comparison, we utilize the same training and testing data as in DeepIPs\cite{lv2021deepips} and adopted the independent test performance reported in these papers. All detailed evaluation metrics related to the independent test data are presented in Table \ref{Table Performance evaluation}. We observe that PTransIPs outperforms the other five predictors. For the S/T sites, PTransIPs achieves the best performance in all five model evaluation metrics (ACC, SEN, SPEC, MCC, AUC), with an AUC value of 0.9232, which is higher by 0.65\%, 3.30\%, 4.12\%, 4.93\%, and 5.36\% compared to DE-MHAIPs, DeepIPs, MusiteDeep2020, MusiteDeep2017, and DeepPSP, respectively. For the Y sites, PTransIPs also performs the best in four out of the five metrics, with ACC of 92.86\% and MCC of 0.8581, outperforming the second best models by 1.60\% and 2.46\% in these two metrics, respectively. We also conducted paired sample t-tests and Wilcoxon signed-rank tests comparing PTransIPs with the previously best method available. The boxplot and results can be found in supplementary material Figure S4. The p-values for S/T sites were 0.0022 and 0.0016, respectively, indicating that the differences are statistically significant. These test results demonstrate that PTransIPs possesses a superior predictive capability compared to the existing tools.


\begin{table*}
\centering
\caption{Independent Testing Performance Comparison between PTransIPs and SOTA Phosphorylation Site Identification Tools for S/T and Y Sites}
\label{Table Performance evaluation}
\begin{tabularx}{\textwidth}{X p{3cm} XXXXXX}
\hline Residue type & Method & Year & ACC & SEN & SPEC & MCC & AUC \\
\hline S/T & PTransIPs &  & \textbf{0.8438} & \textbf{0.8554} & \textbf{0.8323} & \textbf{0.6879} & \textbf{0.9232} \\
& DE-MHAIPs \cite{wang2023DE-MHAIPs} & 2023 & 0.8371 & 0.8428 & 0.8314 & 0.6745 & 0.9172 \\
& DeepIPs \cite{lv2021deepips} & 2021 & 0.8063 & 0.7961 & 0.8350 & 0.6316 & 0.8937 \\
& DeepPSP \cite{guo2021deeppsp} & 2021 & 0.8021 & 0.7665 & 0.8378 & 0.6058 & 0.8762 \\
& MusiteDeep2020 \cite{wang2020musitedeep} & 2020 & 0.8095 & 0.8295 & 0.7896 & 0.6196 & 0.8867 \\
& MusiteDeep2017 \cite{wang2017musitedeep} & 2017 & 0.8017 & 0.7887 & 0.8146 & 0.6035 & 0.8798 \\
\hline
Y & PTransIPs &  & \textbf{0.9286} & \textbf{0.9524} & \textbf{0.9048} & \textbf{0.8581} & 0.9683 \\
& DE-MHAIPs\cite{wang2023DE-MHAIPs} & 2023 & 0.9140 & 0.9507 & 0.8786 & 0.8375 & \textbf{0.9778} \\
& DeepIPs \cite{lv2021deepips} & 2021 & 0.8333 & 0.9048 & 0.8095 & 0.7175 & 0.9252 \\
& DeepPSP \cite{guo2021deeppsp} & 2021 & 0.7619 & 0.9524 & 0.5714 & 0.5665 & 0.8209 \\
& MusiteDeep2020\cite{wang2020musitedeep} & 2020 & 0.8551 & 0.9524 & 0.7619 & 0.7276 & 0.8730 \\
& MusiteDeep2017\cite{wang2017musitedeep} & 2017 & 0.8095 & 0.8571 & 0.7619 & 0.6219 & 0.8141\\
\hline
\end{tabularx}
\end{table*}

\subsection{Adapting PTransIPs for broader applications: a deep learning approach to more bioactivities}

\begin{table*}[ht]
\centering
\caption{Collection of More bioactivities datasets from Publications with SOTA Models and Their Distribution of Positive and Negative Samples}
\label{tab:more_benchmark_datasets}
\begin{tabularx}{\textwidth}{XXXp{2cm}}
\hline 
Bioactivity & Training dataset & Test dataset & Reference \\
\hline 
Blood–Brain Barrier & 100 Positives and 100 negatives & 19 Positives and 19 negatives & \cite{bbppred2021} \\

Anticancer activity (Main dataset) &  689 positives and 689 negatives & 172 positives and 172 negatives & \cite{anticancer2021} \\

Antiviral activity & 2321 Positives and 2321 negatives & 623 Positives and 623 negatives & \cite{antiviral2021} \\
\hline
\end{tabularx}
\end{table*}



\begin{table*}
\centering
\caption{Independent Testing Performance Comparison of PTransIPs and SOTA Phosphorylation Site Identification Tools on Their Corresponding Bioactivity Peptide Datasets}
\label{Table Performance evaluation case study}
\begin{tabularx}{\textwidth}{p{2.5cm}p{2.5cm}XXXXXX}
\hline Bioactivity & Method & Year & ACC & SEN & SPEC & MCC & AUC \\
\hline 
Blood–Brain Barrier
& PTransIPs &  & \textbf{0.8947} & 0.8421 & \textbf{0.9474} & \textbf{0.7939} & 0.9418 \\
& UniDL4BioPep\cite{du2023unidl4biopep} & 2023 & 0.842 & \textbf{0.882} & 0.809 & 0.688 & \textbf{0.992} \\
& BBPpred\cite{bbppred2021} & 2021 & 0.7895 & 0.6316 & 0.9474 & 0.6102 & 0.7895 \\
\hline
Anticancer activity
& PTransIPs &  & 0.7442 & \textbf{0.8488} & 0.6395 & 0.4994 & \textbf{0.8505} \\
& UniDL4BioPep\cite{du2023unidl4biopep} & 2023 & 0.735 & 0.734 & 0.737 & 0.471 & 0.805 \\
& iACP-FSCM\cite{anticancer2021} & 2021 & \textbf{0.825} & 0.726 & \textbf{0.903} & \textbf{0.646} & 0.81 \\
\hline
Antiviral activity
& PTransIPs &  & \textbf{0.8515} & 0.8202 & 0.8828 & \textbf{0.7044} & \textbf{0.9236} \\
& UniDL4BioPep\cite{du2023unidl4biopep} & 2023 & 0.842 & \textbf{0.916} & 0.79 & 0.694 & 0.907 \\
& ABPDiscover\cite{antiviral2021} & 2021 & 0.828 & 0.764 & \textbf{0.892} & 0.662 & 0.896 \\

\hline
\end{tabularx}
\end{table*}

To evaluate the generalization ability of PTransIPs across various bioactivities, particularly within peptide datasets, we have sourced and analyzed several datasets from state-of-the-art (SOTA) models. 
We obtained all the benchmark datasets from models documented by \cite{du2023unidl4biopep}, ensuring a balanced and impartial performance analysis. Here we adapt PTransIPs for three different bioactivities of very different data size, including Blood–Brain Barrier \cite{bbppred2021}, anticancer activity \cite{anticancer2021}, and antiviral activity \cite{antiviral2021}. The specifics of each dataset are presented in Table \ref{tab:more_benchmark_datasets}, with in-depth descriptions available in earlier publications.

We train our model using 5-fold cross-validation for each dataset mentioned. To ensure fairness and maintain consistency, all hyperparameters are kept identical to the training on phosphorylation sites. Notably, the sequences in these datasets differ significantly in length compared to the uniformly lengthed phosphosites used previously. Therefore, to maintain stability in the training process, we introduce necessary modifications such as padding in the encoding function of sequences and omitting structure pre-trained embedding.

The performance results suggest that PTransIPs consistently has high accuracy across all the bioactivity datasets in this part, shown in Table \ref{Table Performance evaluation case study}. Particularly, for Blood-Brain Barrier activity prediction, PTransIPs achieves superior ACC of 0.8947 and MCC of 0.7939 in comparison with UniDL4BioPep\cite{du2023unidl4biopep} and BBPpred\cite{bbppred2021}; for anticancer activity, PTransIPs performs better on AUC of 0.8505 comparison with UniDL4BioPep and iACP-FSCM\cite{anticancer2021}; for antiviral activity, PTransIPs outperforms both UniDL4BioPep and ABPDiscover\cite{antiviral2021} in terms of ACC of 0.8515, MCC of 0.7044 and AUC of 0.9236.

These results show that PTransIPs not only possesses the ability to identify phosphorylation sites, but also holds the potential as a reliable model for predicting peptide datasets associated with various bioactivities.

\section{Discussion}\label{Discussion}

The core innovation of PTransIPs lies in its use of embeddings generated by pre-trained protein models as additional information beyond phosphorylation peptide sequence data, thereby enhancing model performance. This approach helps the model converge more effectively to a global optimum and demonstrates significant generalization capabilities. Ablation experiments conducted in our study have shown that incorporating features extracted from pre-trained models as additional inputs can enhance the model's overall predictive performance. Moreover, visualizing these extracted features using the UMAP method, we found that they inherently have discriminative capabilities to distinguish between different types of data, indicating the effectiveness of the PTransIPs architecture.

The limitations and future works of PTransIPs are mainly in two aspects. The first is about the selection of protein PLMs. In this paper, we specifically use two existing protein PLMs ProtTrans and EMBER2 to pre-extract sequence and structural features, and then integrate them into the training process to enhance features, thus achieving excellent performance. Given the rapid progress in the AI field, using embeddings generated by new protein pre-trained models in the future could further improve performance for this task. The second limitation is the challenge brought by dataset imbalance and unequal peptide lengths for model training. In the generalization experiments of PTransIPs, we adapt PTransIPs to more complex peptide datasets. We handle peptides of varying lengths by padding them to a uniform length, but this is a relatively crude approach. There are also some overly short peptides in the dataset, which may reduce the effectiveness of embeddings generated by protein pre-trained models. Considering that we did not achieve the best performance on all metrics in the extended tasks, using more effective data augmentation or new approaches such as graph neural networks might be potential ways to improve the performance of such models.

\section{Conclusion}\label{Conclusion}

In this study, we introduce PTransIPs, a deep learning model for identifying phosphorylation sites. Independent tests have demonstrated that for recognizing phosphorylated S/T and Y sites, PTransIPs achieved AUCs of 0.9232 and 0.9660, respectively, surpassing other existing models. Moreover, PTransIPs can be generalized to other bioactivity classification tasks, maintaining performance on par with state-of-the-art models.

PTransIPs contributes in the following three main aspects: 1) Enhancing model performance using embeddings generated by protein pre-trained language models, with its effectiveness proven through ablation studies and intuitively explained through UMAP visualization. 2) Achieving superior performance using Transformer-based models and TIM loss function, providing practical experience. 3) Serving as a universal framework adaptable to any peptide-based bioactivity task, highlighting its remarkable generalization capability. Additionally, we have made our source code and data access available at \url{https://github.com/StatXzy7/PTransIPs}.

In conclusion, our research results demonstrate that PTransIPs can effectively identify phosphorylation sites. We hope that PTransIPs will serve as a powerful tool, contributing to a deeper understanding of phosphorylation sites and other bioactivities.




\section*{References}

\bibliographystyle{ieeetr}
\bibliography{reference}

\end{document}